\documentclass[aps,twocolumn,showpacs,groupedaddress]{revtex4}
\usepackage{graphicx}

\begin{document}
\def\dbarit {{\mathchar'26\mkern-11mud}}
\title{Quantum heat engine with continuum working medium }
\author{S. Li, H. Wang, Y. D. Sun, and
X. X. Yi} \affiliation{Department of Physics, Dalian University of
Technology, Dalian 116024, China }

\date{\today}

\begin{abstract}
We introduce a new quantum heat engine, in which the working
medium is a quantum system with a discrete level and a continuum.
Net work done by this engine is calculated and discussed. The
results show that this quantum heat engine behaves like the
two-level quantum heat engine in  both the high-temperature and
the low-temperature  limits, but it operates differently in
temperatures between them. The efficiency of this quantum heat
engine is also presented and discussed.
\end{abstract}

 \maketitle

A classical heat engine converts heat energy into mechanical work
by using a classical-mechanical system in which a working
medium(for example, a gas) expands and pushes a piston in a
cylinder. Working between a high-temperature reservoir and a
low-temperature reservoir, the classical heat engine achieves
maximum efficiency when it is reversible, while the efficiency is
zero if the two reservoirs have the same temperature. The
situation changes for its quantum counterpart, where the working
medium and the dynamics that govern the cycle are quantum. It was
shown that the quantum heat engine can better the work extraction
and improve the engine efficiency\cite{scully03, kieu03, kieu04,
quan05, quan05p}.

The quantum heat engine concept was introduced by Scovil and
Schultz-Dubois \cite{scovil59} and extended in many later
works\cite{lloyd97,kosloff00,scully03, kieu03, kieu04, quan05,
quan05p}. Quantum heat engines are characterized by three
attributes: the working medium, the cycle of operation, and the
dynamics that govern the cycle. In the previous works, the working
medium is considered as an ensemble of many non-interacting
discrete level systems. Specifically, the analysis is carried out
on two-level systems\cite{kieu03, feldmann00}, three-level
systems\cite{quan05}, as well as an ensemble of harmonic
oscillators\cite{kieu03, feldmann96, bender01}. These give rise to
the following questions, with continuum working medium what is the
work extraction of quantum heat engine? Can such a quantum heat
engine improve the work extraction?

In this paper we will answer these questions by examining an
quantum heat engine working between two reservoirs with different
temperatures. The working medium is envisioned as a quantum system
with a discrete level $|d\rangle$ and a continuum $|c\rangle$ as
shown in figure 1.
 \begin{figure}
\includegraphics*[width=0.8\columnwidth,
height=0.4\columnwidth]{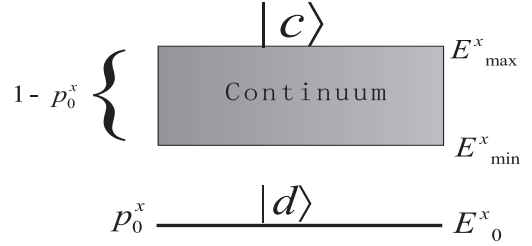} \caption{An illustration of the
level structure of the working medium. The occupation probability
$p_0^x, x=l,h$ of the discrete level $|d\rangle$ (with eigenenergy
$E_0^x$) was kept fixed in adiabatic processes. The continuum
broadening was denoted by $E_{max}^x-E_{min}^x$ . } \label{fig1}
\end{figure}
Non-interacting Ba atoms, for example, may serve as the quantum
system\cite{noordam92}, in which the bound coherent Rydberg state
$6s(n+2)d$ might be taken as the discrete level. The heat-engine
cycle consists of four branches labelled by 1,2,3 and 4, this was
schematically illustrated in figure 2.
\begin{figure}
\includegraphics*[width=0.9\columnwidth,
height=0.8\columnwidth]{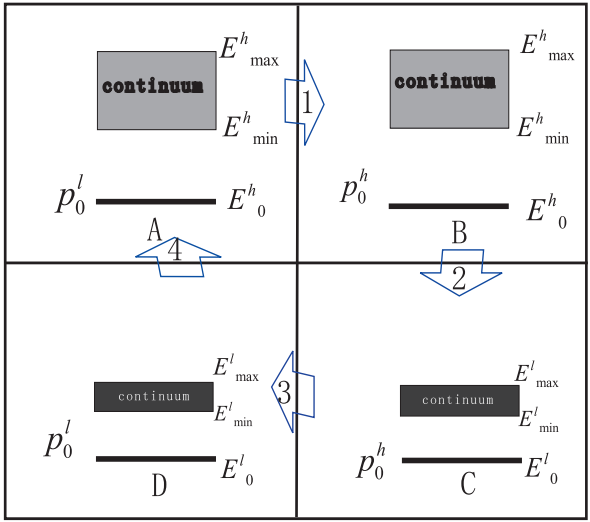} \caption{ Schematic
illustration of the four-stroke quantum heat engine. From state A
to B, the working medium absorbs  heat from the high-temperature
reservoir, leading to population transfer from the discrete level
$|d\rangle$ to the continuum. From state B to C, works are done
with the working medium undergoing an adiabatic process. The
Branches 3 and 4(corresponding changes from C to D, and from D to
A, respectively.) are reversed processes of 1 and 2, respectively.
We will use $(E_0^x, E_{min}^x, E_{max}^x)$  to characterize the
level structure of the working medium in the text.} \label{fig2}
\end{figure}
This four-stroke quantum heat engine is  a quantum analogue of the
classical Otto engine, which includes two quantum adiabatic
processes (2 and 4) and two isothermal processes (1 and 3).

Denoting $p_0^x,$ $ {x=h,l}$ the occupation probability of the
discrete level and $p(E)$ the occupation probability of the
continuum, we can write the expectation value of the measured
energy of a quantum system $U$ as,
\begin{equation}
U=\langle E\rangle=\sum_i p_iE_i+ \int d[p(E)\cdot E].
\end{equation}
The definition of infinitesimal work done in a process is then
\begin{equation}
\dbarit W=\sum_i p_i dE_i+\sum^{E_{max}}_{E_{min}}
p(E)dE,\label{work}
\end{equation}
which is a straightforward extension of that for discrete level
systems\cite{feldmann00} to the system under our consideration.
The first term comes from the contribution of discrete levels,
while the last term comes from the continuum. By the first law of
thermodynamics $dU=\dbarit Q +\dbarit W$, the infinitesimal heat
absorbed is
\begin{equation}
\dbarit Q=\sum_i E_i dp_i+\sum^{E_{max}}_{E_{min}} E
dp(E).\label{heat}
\end{equation}
With these notations, we now calculate the work done on the four
branches of the heat engine.

In the branch 1, namely, from A to B, the working medium is
coupled to a hot reservoir of temperature $T_h$ and its energy
structure is kept fixed. In this isothermal process, the
population of the discrete level is changing from the initial
population $p_0^l$ to the population $p_0^h$. Accordingly, the
total population of the continuum is changing from $1-p_0^l$ to
$1-p_0^h$. The work done in this branch clearly is  zero by the
definition Eq.(\ref{work}). In the branch 2, B $\rightarrow$ C,
the working medium is decoupled from the hot reservoir, and the
energy structure is varied from $(E_0^h, E_{min}^h, E_{max}^h)$ to
$(E_0^l, E_{min}^l, E_{max}^l).$ In this process, the occupation
probability $p_0^h$ is kept fixed. This is an adiabatic process in
the sense that the total occupation probability of the working
medium on the continuum remains unchanged, but population transfer
among states in the continuum  is allowed. Strictly speaking, the
evolution of the working medium in this branch is not adiabatic.
However, it may be seen as an adiabatic one provided the continuum
is treated as an energy level. After the working medium reaches
thermodynamical equilibrium, the total occupation probabilities of
the working medium on the continuum satisfy (coming from
equilibrium state B, C, D and A, respectively),
\begin{eqnarray}
1-p_{0}^{h}&=&\int_{E_{min}^{h}}^{E_{max}^{h}}
\frac{\rho_{h}}{Z_{hh}}e^{-\beta_{h}E^{h}}dE^{h},  \nonumber\\
1-p_{0}^{h}&=&\int_{E_{min}^{l}}^{E_{max}^{l}}
\frac{\rho_{l}}{Z_{hl}}e^{-\beta_{h}E^{l}}dE^{l}\nonumber\\
&=&\int_{E_{min}^{h}}^{E_{max}^{h}}\frac{\rho_{h}}{Z_{hl}}
e^{-\beta_{h}(\frac{\rho_{h}}
{\rho_{l}}(E^{h}-E^{h}_{min})+E^{l}_{min})}dE^h,  \nonumber\\
1-p_{0}^{l}&=&\int_{E_{min}^{l}}^{E_{max}^{l}}
\frac{\rho_{l}}{Z_{ll}}e^{-\beta_{l}E^{l}}dE^{l}\nonumber\\
&=&\int_{E_{min}^{h}}^{E_{max}^{h}}\frac{\rho_{h}}{Z_{ll}}
e^{-\beta_{l}(\frac{\rho_{h}}{\rho_{l}}(E^{h}-E^{h}_{min})
+E^{l}_{min})}dE^h,  \nonumber\\
1-p_{0}^{l}&=&\int_{E_{min}^{h}}^{E_{max}^{h}}
\frac{\rho_{h}}{Z_{lh}}e^{-\beta_{l}E^{h}}dE^{h}.\label{limit1}
\end{eqnarray}
Here $\rho_h$ ($ \rho_l$) denotes the degeneracy  of the continuum
with level structure $(E_0^h, E_{min}^h, E_{max}^h)$ ($(E_0^l,
E_{min}^l, E_{max}^l)$), and it is assumed to be constant. $E^x,
x=h,l$ stand for an eigenenergy in the continuum with level
structure $(E_0^x, E_{min}^x, E_{max}^x)$.   $Z_{hh}, Z_{hl},
Z_{ll}$ and $Z_{lh}$ are the partition function of the working
medium at equilibrium state B, C, D and A, respectively. $
\beta_{h}=\frac{1}{KT_{h}},$ $ \beta_{l}=\frac{1}{KT_{l}},$ and
$K$ is the Boltzmann constant.
  Throughout this paper, we focus our attention on
the following situation,
\begin{eqnarray}
\rho_{h}(E_{max}^{h}-E_{min}^{h})&=&\rho_{l}
(E_{max}^{l}-E_{min}^{l}),\nonumber\\
\rho_{h}(E^{h}-E_{min}^{h})&=&\rho_{l}(E^{l}-E_{min}^{l}).
\end{eqnarray}
These relations mean that the adiabatic process does not change
the distribution of microstates. In other words, the degeneracy of
the continuum is supposed to be changed homogeneously in adiabatic
processes. In fact, this relation was used in 3rd and 5th lines in
Eq. (\ref{limit1}).

Define,
\begin{eqnarray}
E_{min}^{h}-E_{0}^{h}=\Delta^{h}\nonumber,
E_{max}^{h}-E_{min}^{h}=\delta^{h}\nonumber\\
E_{min}^{l}-E_{0}^{l}=\Delta^{l}\nonumber,
E_{max}^{l}-E_{min}^{l}=\delta^{l}.
\end{eqnarray}
The energy change in the branch 2 reads,
\begin{widetext}
\begin{eqnarray}
\Delta U_2&=&p_{0}^{h}(E_{0}^{h}-E_{0}^{l})+
\int_{E_{min}^{h}}^{E_{max}^{h}}
\frac{\rho_{h}}{Z_{hh}}e^{-\beta_{h}E^{h}}E^{h}dE^{h}
-\int_{E_{min}^{l}}^{E_{max}^{l}}
\frac{\rho_{l}}{Z_{hl}}e^{-\beta_{h}E^{l}}E^{l}dE^{l}\nonumber\\
&=&p_{0}^{h}(E_{0}^{h}-E_{0}^{l})+\rho_{h}
\int_{E_{min}^{h}}^{E_{max}^{h}}\frac{1}{Z_{hh}}
e^{-\beta_{h}E^{h}}(E^{h}-(\frac{\rho_{h}}
{\rho_{l}}(E^{h}-E_{min}^{h})+E_{min}^{l}))dE^{h}\nonumber\\
&+&\rho_{h}\int_{E_{min}^{h}}^{E_{max}^{h}}
(\frac{1}{Z_{hh}}e^{-\beta_{h}E^{h}}
-\frac{1}{Z_{hl}}e^{-\beta_{h}(\frac{\rho_{h}}
{\rho_{l}}(E^{h}-E_{min}^{h})+E_{min}^{l})})
(\frac{\rho_{h}}{\rho{l}}(E^{h}-E_{min}^{h})
+E_{min}^{l})dE^{h}.\label{u2}
\end{eqnarray}

According to Eq.(\ref{work}), the work done in this branch reads,
\begin{eqnarray}
\Delta W_{2}&=&p_{0}^{h}(E_{0}^{h}-E_{0}^{l})
+\int_{E_{min}^{h}}^{E_{max}^{h}}\frac{\rho_{h}}{Z_{hh}}e^{-\beta_{h}E^{h}}
(E^{h}-(\frac{\rho_{h}}{\rho_{l}}(E^{h}-E_{min}^{h})
+E_{min}^{l}))dE^{h}\nonumber\\
&=&p_{0}^{h}(E_{0}^{h}-E_{0}^{l})
+(1-p_{0}^{h})\frac{\rho_{h}E_{min}^{h}-\rho_{l}E_{min}^{l}}{\rho_{l}}
+(1-p_{0}^{h})(\frac{\rho_{l}-\rho_{h}}{\rho_{l}}
(E_{max}^{h}+\frac{\delta^{h}}
{e^{-\beta_{h}\delta^{h}}-1})+\frac{\rho_{l}-\rho_{h}}{\beta_{h}\rho_{l}}),
\end{eqnarray}
\end{widetext}
 which is exactly the second line in Eq.(\ref{u2}). The branch
3 is similar to the first. The working medium is now coupled to a
cold reservoir at temperature $T_l$ and its energy structure is
kept fixed. The occupation probability backs on this branch from
$p_0^h$ to $p_0^l$. The branch 4 closes the cycle and is similar
to the branch 2. The working medium is decoupled from the cold
reservoir, and the level structure is changed  back to its
original value $(E_0^h, E_{min}^h, E_{max}^h)$. Similar analysis
shows that the work done in the  branch 3 is zero, whereas the
energy change and the work done in the branch 4 are
\begin{widetext}
\begin{eqnarray}
-\Delta
U_{4}&=&p_{0}^{h}(E_{0}^{h}-E_{0}^{l})
+\int_{E_{min}^{h}}^{E_{max}^{h}}
\frac{\rho_{h}}{Z_{lh}}e^{-\beta_{l}E^{h}}E^{h}dE^{h}
-\int_{E_{min}^{l}}^{E_{max}^{l}}
\frac{\rho_{l}}{Z_{ll}}e^{-\beta_{l}E^{l}}E^{l}dE^{l}\nonumber\\
&=&p_{0}^{l}(E_{0}^{h}-E_{0}^{l})
+\rho_{h}\int_{E_{min}^{h}}^{E_{max}^{h}}
(\frac{1}{Z_{lh}}e^{-\beta_{l}E^{h}}-\frac{1}{Z_{ll}}e^{-\beta_{l}
(\frac{\rho_{h}}{\rho_{l}}(E^{h}
-E_{min}^{h})+E_{min}^{l})})E^{h}dE^{h}\nonumber\\
&+&\rho_{h}\int_{E_{min}^{h}}^{E_{max}^{h}}
\frac{1}{Z_{ll}}e^{-\beta_{l}
(\frac{\rho_{h}}{\rho_{l}}(E^{h}-E_{min}^{h})
+E_{min}^{l})}(E^{h}-(\frac{\rho_{h}}{\rho_{l}}
(E^{h}-E_{min}^{h})+E_{min}^{l}))dE^{h},
\end{eqnarray}

\begin{eqnarray}
-\Delta W_
{4}&=&p_{0}^{l}(E_{0}^{h}-E_{0}^{l})+\rho_{h}
\int^{E^{h}_{max}}_{E^{h}_{min}}\frac{1}{Z_{ll}}
e^{-\beta_{l}(\frac{\rho_{h}}{\rho_{l}}(E^{h}-E^{h}_{min})
+E^{l}_{min})}(E^{h}-(\frac{\rho_{h}}{\rho_{l}}(E^{h}-E^{h}_{min})
+E^{l}_{min}))dE^{h}\nonumber\\
&=&p_{0}^{l}(E_{0}^{h}-E_{0}^{l})+(1-p_{0}^{l})
\frac{\rho_{h}E^{h}_{min}-\rho_{l}E^{l}_{min}}
{\rho_{l}}+(1-p_{0}^{l})(\frac{\rho_{l}-\rho_{h}}
{\rho_{l}}(E^{h}_{max}+\frac{\delta^{h}}{e^{-\beta_{l}
\frac{\rho_{h}}{\rho_{l}}\delta^{h}}-1})+\frac{\rho_{l}-
\rho_{h}}{\beta_{l}\rho_{h}}).
\end{eqnarray}
The net work done in the whole cycle is then
\begin{eqnarray}
\Delta W&=&\Delta W_{2}+\Delta
W_{4}=(p_{0}^{h}-p_{0}^{l})(E_{0}^{h}-E_{0}^{l})+
(p_{0}^{l}-p_{0}^{h})\frac{\rho_{h}E^{h}_{min}
-\rho_{l}E^{l}_{min}}{\rho_{l}}+
(p_{0}^{l}-p_{0}^{h})\frac{\rho_{l}-\rho_{h}}
{\rho_{l}}E^{h}_{max}\nonumber\\&+&\frac{\rho_{l}
-\rho_{h}}{\rho_{l}}\delta^{h}(\frac{1-p_{0}^{h}}
{e^{-\beta_{h}\delta^{h}}-1}
-\frac{1-p_{0}^{l}}{e^{-\beta_{l}\frac{\rho_{h}}
{\rho_{l}}\delta^{h}}-1})+(\rho_{l}-\rho_{h})
(\frac{1-p_{0}^{h}}{\beta_{h}\rho_{l}}
-\frac{1-p^{l}_{0}}{\beta_{l}\rho_{h}}).
\end{eqnarray}
\end{widetext}
Noticing
\begin{eqnarray}
\frac{\rho_{l}}{\rho_{h}}=\frac{\delta^{h}}{\delta^{l}},
\end{eqnarray}
one can reduce the net work $\Delta W$ to
\begin{eqnarray}
\Delta W&=&(p^{l}_{0}-p^{h}_{0})((\Delta^{h}+\delta^{h})
-(\Delta^{l}+\delta^{l}))\nonumber\\
&+&( \delta^{h}-\delta^{l})\cdot f(p_0^x,T_x,\delta^x)|_{x=l,h}.\nonumber\\
\label{cr1}
\end{eqnarray}
Here
\begin{eqnarray}
f(p_0^x,T_x,\delta^x)|_{x=l,h}&=&\frac{1-p_{0}^{h}}
{e^{-\frac{\delta^{h}}{KT_{h}}}-1}-\frac{1-p^{l}_{0}}
{e^{-\frac{\delta^{l}}{KT_{l}}}-1} +\nonumber\\
\frac{1-p^{h}_{0}}{\frac{\delta^{h}}{KT_{h}}}
-\frac{1-p^{l}_{0}}{\frac{\delta^{l}}{KT_{l}}}.
\end{eqnarray}
This is the central result of this paper, showing that the net
work done by the heat engine depends on the occupation
probabilities $p_0^h$ and $p_0^l$, the continuum broadenings
$\delta^h$ and $\delta^l$, the energy gaps $\Delta^h$ and
$\Delta^l$ as well as the low  and high temperature of the
reservoir. To get more insight in this result, we consider the
following limiting situations. (a) The continuum broadening
remains unchanged in the cycle, namely, $\delta^{h}=\delta^{l}$.
The net work in this case reads,
\begin{eqnarray}
\Delta
W_{\delta}=(p^{l}_{0}-p^{h}_{0})(\Delta^{h}-\Delta^{l}).\label{rework1}
\end{eqnarray}
This backs to the net work done by the quantum heat engine with
two-level systems as its working medium. (b)High temperature
limit, $\frac{\delta^{h}}{KT_{h}}\ll1,
\frac{\delta^{l}}{KT_{l}}\ll1.$ The net work done in this
situation follows
\begin{eqnarray}
\Delta
W_T&=&(p^{l}_{0}-p^{h}_{0})((\Delta^{h}+\delta^{h})-(\Delta^{l}+\delta^{l})).
\end{eqnarray}
Interestingly, the net work   in this case takes the same form as
in Eq.(\ref{rework1}), but the energy difference of the two-level
working medium  is $(\Delta^h+\delta^h)$ at high temperature and
$(\Delta^l+\delta^l)$ at low temperature. The same results are
found in low temperature limit. (c) No population transfer between
the discrete level and the continuum in the cycle, i.e.,
$p_0^h=p_0^l=p$. The net work $\Delta W$ in this case follows from
Eq.(\ref{cr1}),
\begin{eqnarray}
\Delta W_p=( \delta^{h}-\delta^{l})(1-p)\cdot
f(p_0^x=0,T_x,\delta^x)|_{x=l,h} .
\end{eqnarray}
As shown, $\Delta W_p$ totally comes from the contribution of the
continuum. It is zero if $\delta^h=\delta^l$, and it increases
linearly with $p$ decreases. $\Delta W_p>0$ requires that
$T_h>\frac{\delta^h}{\delta^l}T_l$ and $\delta^h>\delta^l.$ This
 is similar to the requirement upon the two-level quantum heat
engine\cite{geva92,scully03} for positive work
 extraction. In order to compare our heat engine
with the two-level one, we plot a work difference $(\Delta
W-\Delta W_{\delta})$ versus $\delta^h$ and $\delta^l$ in figure
3.  Note that this work difference is different from $\Delta W_p$,
where $p_0^l=p_0^h=p$ is considered. As we mentioned above, the
contribution from the continuum was excluded in $\Delta
W_{\delta}$. So, $(\Delta W-\Delta W_{\delta})$ mostly
characterize the effect of the continuum on the work extraction in
the quantum heat engine. From the other aspect, this work
difference can be understood as the net work with the energy gap
unchanged in the cycle, i.e., $\Delta^l=\Delta^h.$  Figure 3 shows
that the work difference decreases with $\delta^l$ increases for
small $\delta^h$, but the result goes in the opposite direction
for large $\delta^h$. We also find from figure 3 that $(\Delta
W-\Delta W_{\delta})>0$ in the region $\delta^l\simeq \delta^h$
and around.
\begin{figure}
\includegraphics*[width=0.8\columnwidth,
height=0.65\columnwidth]{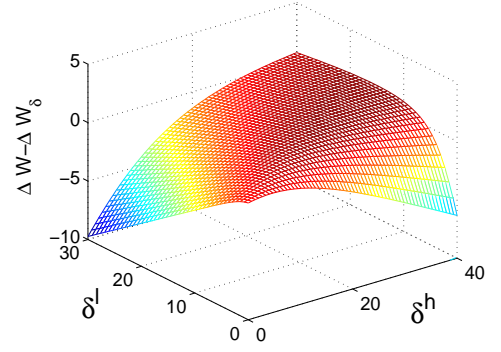} \caption{The   work difference
($\Delta W-\Delta W_{\delta}$) as a function of $\delta^h$ and
$\delta^l$. The parameters chosen are $p_0^l=0.5$, $p_0^h=0.3$,
and $KT_h=5$. The work difference, $\delta^x,(x=l,h)$ and $KT_h$
were rescaled in units of $KT_l=1$ in this plot. } \label{fig3}
\end{figure}

Now, we turn to study the efficiency of this quantum heat engine.
It is defined as the ratio of the work done to the heat absorbed
in the cycle,
\begin{equation}
\eta=\frac{\Delta W}{\Delta Q}.
\end{equation}
By the first law of thermodynamics, we have
\begin{equation}
\eta=1-\frac{\Delta Q_{2}+\Delta Q_{3}}{\Delta Q_{1}+\Delta
Q_{4}},
\end{equation}
where $\Delta Q_i (i=1,2,3,4)$ denote the heat exchange between
the working medium and reservoirs on the branch $i$. By the same
procedure presented above, one obtains
\begin{eqnarray}
\Delta Q_{1}+\Delta
Q_{4}&=&(p_0^l-p_0^h)(\Delta^h+\delta^h)\nonumber\\
&+&\delta^h \cdot f(p_0^x,T_x,\delta^x)|_{x=l,h},\nonumber\\
\Delta Q_{2}+\Delta
Q_{3}&=&(p^{l}_{0}-p^{h}_{0})(\Delta^l+\delta^l)\nonumber\\
&+&\delta^l\cdot f(p_0^x,T_x,\delta^x)|_{x=l,h}.
\end{eqnarray}
Then the efficiency of the quantum heat engine reads,
\begin{eqnarray}
\eta=1-\frac{(p^{l}_{0}-p^{h}_{0}) (\Delta^l+\delta^l)+ \delta^l
\cdot f(p_0^x,T_x,\delta^x)|_{x=l,h}} {(p^{l}_{0}-p^{h}_{0})
(\Delta^h+\delta^h)+ \delta^h \cdot
f(p_0^x,T_x,\delta^x)|_{x=l,h}}.\nonumber\\
\end{eqnarray}
In the high-temperature limit $\frac{\delta^{h}}{KT_{h}}\ll1,
\frac{\delta^{l}}{KT_{l}}\ll1,$ $\eta$ reduces to
\begin{eqnarray}
\eta=1-\frac{\Delta^l+\delta^l}{\Delta^h+\delta^h},
\end{eqnarray}
returning back to the efficiency of the two-level quantum heat
engine. This observation holds in the low-temperature, as the net
work does. Similarly, for $p_0^l=p_0^h$, the efficiency becomes
$\eta=1-\delta^l/\delta^h.$ Note that in the limit
$\delta^l=\delta^h$, the net work $\Delta W$ returns back to the
result of the two-level quantum heat engine, but the efficiency
does not. This is due to the difference in heat exchange of the
two engines.

 In conclusion, a new quantum heat engine has been introduced in this
paper. As its working medium, the quantum system has a discrete
level and a continuum. This makes the engine different from the
two-level quantum heat engine. The quantum heat engine consists of
two adiabatic processes and two isothermal processes. It can
extract work like a two-level quantum heat engine in the
high-temperature and low-temperature limits, whereas it works in a
different way at temperatures between the two. Since the previous
studies on quantum heat engine were focused on various working
mediums only with discrete energy levels, the study presented here
can better the understanding of quantum heat engine, and might
provides  us a new way to study the unsolved problems of emergence
of classicality.

\ \ \\
 This work was supported by EYTP of M.O.E,  NSF of China
(10305002 and 60578014).

\end{document}